\documentclass[11pt]{article}
\usepackage{graphicx}
\usepackage{url}
\usepackage{makecell}
\usepackage{xcolor}
\usepackage{framed}
\usepackage{caption}
\usepackage{subcaption}
\usepackage{float}
\usepackage{tcolorbox}
\usepackage{textcomp}
\usepackage{hyperref}
\usepackage{multicol}
\usepackage{blindtext}
\usepackage{amssymb}
\usepackage{colortbl}
\usepackage{booktabs} 
\usepackage[colorinlistoftodos]{todonotes}
\usepackage{geometry}
\usepackage{epigraph}
 \let\OLDthebibliography\thebibliography
\renewcommand\thebibliography[1]{
  \OLDthebibliography{#1}
  \setlength{\parskip}{0pt}
  \setlength{\itemsep}{0pt plus 0.5ex}
}
 
\definecolor{lightgray}{gray}{0.94}

\newcounter{mainfindingid}
\newcommand{\mainfinding}[2]{\refstepcounter{mainfindingid}\label{#1}\item[
	\ifthenelse{\value{mainfindingid}=1}{Perspective-}{P-}\arabic{mainfindingid}:
	] #2}

\newcounter{recommendationid}
\newcommand{\recommendation}[2]{\refstepcounter{recommendationid}\label{#1}\item[
	\ifthenelse{\value{recommendationid}=1}{Recommendation-}{R-}\arabic{recommendationid}:
	] #2}

\usepackage{hyperref}
\hypersetup{
    colorlinks=true,
    linkcolor=black,
    filecolor=red,      
    urlcolor=red,
    citecolor=blue
}

\newtcolorbox{mybox}[2][]{
  colframe = #2!25,
  colback  = #2!10,
  #1,
}

\usepackage[format=plain,
            labelfont={bf,it},
            textfont={bf,it}]{caption}

\begin{document}

\thispagestyle{empty}

\begin{center}

Vrije Universiteit Amsterdam

\vspace{1mm}

\includegraphics[height=28mm]{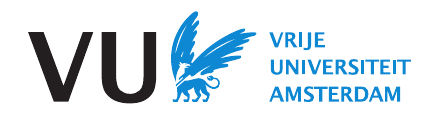}

\vspace{1.5cm}

{\Large Master Thesis}

\vspace*{1.5cm}

\rule{.9\linewidth}{.6pt}\\[0.4cm]
{\huge \bfseries Evaluating the Effects of Digital Privacy Regulations on User Trust
 \par}
\rule{.9\linewidth}{.6pt}\\[1.5cm]

\vspace*{2mm}

{\Large
\begin{tabular}{l}
{\bf Author:} ~~Mehmet Berk Cetin~~~~ (2644886)
\end{tabular}
}

\vspace*{1.5cm}

\begin{tabular}{ll}
{\it 1st supervisor:}   & ~~Dr. Anna Bon \\
{\it 2nd reader:}   & ~~Dr. Hans Akkermans \\

\end{tabular}

\vspace*{1cm}

\today\\[4cm] 
\end{center}

\pagenumbering{roman}

\newpage
\vfill
\noindent\makebox[\linewidth]{\rule{\textwidth}{0.4pt}}
\begin{quote}
    \centering
    \textit{``By three methods we may learn wisdom: First, by reflection, which is noblest; Second, by imitation, which is easiest; and third by experience, which is the bitterest.''} \\
    \textit{by Confucius}
\end{quote}
\vfill
\newpage

\renewcommand*\contentsname{Table of Contents\newline}
\tableofcontents
\newpage

\begin{abstract}
In today's digital society, issues related to digital privacy have become increasingly important. Issues such as data breaches result in misuse of data, financial loss, and cyberbullying, which leads to less user trust in digital services. This research investigates the impact of digital privacy laws on user trust by comparing the regulations in the Netherlands, Ghana, and Malaysia. The study employs a comparative case study method, involving interviews with digital privacy law experts, IT educators, and consumers from each country. The main findings reveal that while the General Data Protection Regulation (GDPR) in the Netherlands is strict, its practical impact is limited by enforcement challenges. In Ghana, the Data Protection Act is underutilized due to low public awareness and insufficient enforcement, leading to reliance on personal protective measures. In Malaysia, trust in digital services is largely dependent on the security practices of individual platforms rather than the Personal Data Protection Act. The study highlights the importance of public awareness, effective enforcement, and cultural considerations in shaping the effectiveness of digital privacy laws. Based on these insights, a recommendation framework is proposed to enhance digital privacy practices, also aiming to provide valuable guidance for policymakers, businesses, and citizens in navigating the challenges of digitalization.

\end{abstract}

\newpage

\section{Introduction} \label{sec:introduction} \pagenumbering{arabic} 

Imagine waking up one morning to discover that your personal data, consisting of your name, email, and even credit card information, has been compromised in a massive data breach. This unsettling scenario recently affected millions of users of the app MyFitnessPal, where a data breach exposed the personal information of over 150 million users, leading to widespread concerns about data security and privacy. Users' email addresses, usernames, and hashed passwords were among the compromised data, highlighting the risks associated with sharing personal information online without fully understanding how it will be protected~\cite{dickey2018under}.
Despite such incidents, users of online digital services frequently feel obligated to share their personal data to utilize various services, yet they often disregard the terms and conditions that explain how their data will be used \cite{usatodayWhatNeed}, or they find these terms simply too complicated to understand \cite{van2019does}. This lack of clarity and level of complexity can result in personal data being sold or compromised in a data breach without the user's awareness \cite{steel2010facebook}. Moreover, the online exposure of personal information can lead to its abuse \cite{moore2009economics, thomas2006underground}, financial damage \cite{abbasi2010detecting}, and cyberbullying \cite{juvonen2008bully}. Such incidents demolish user trust \cite{union2008consumer} and negatively affect the digital economy \cite{moore2009economics}. Therefore, safeguarding digital privacy is crucial given the severe consequences data breaches have on consumers.

To tackle various privacy and security issues, countries and regions have implemented their own regulations. Examples include China's Personal Information Protection Law (PIPL) enacted in 2021 \cite{calzada2022citizens}, the European Union's General Data Protection Regulation (GDPR) introduced in 2018 \cite{gdprWhatGDPR}, Malaysia's Personal Data Protection Law (PDPL) from 2010 \cite{malaysiaPrivacyAct}, the California Consumer Privacy Act (CCPA) established in the United States \cite{caCaliforniaConsumer}, and many more around the world. While there are many privacy and security laws worldwide, the GDPR is recognized as the strictest \cite{gdprWhatGDPR, Krishnamurthy_2020}. Companies handling data from EU citizens must comply with GDPR regulations or face fines of up to tens of millions of euros \cite{gdprWhatGDPR}. GDPR empowers consumers with greater control over their data, including rights to withdraw consent or request data deletion \cite{li2019gdprimpact}.
Different privacy laws have distinct focuses. For example, GDPR emphasizes the nationality of the data subject and the location of the business, while PIPL focuses on where the data processing occurs \cite{chinabriefingPIPLGDPR}. 

\subsection{Background on digital privacy regulations} \label{sec:problem}
Digital privacy laws can affect business practices \cite{quach2022digital} and user behavior \cite{schmitt2020impact} when consuming digital services. Therefore, to have better digital services that are safer, more private, and hence more attractive to consumers, understanding the impact of privacy regulations on digital service consumption is critical in today's globalized digital landscape. 
To explore this impact, we select three different countries, namely the Netherlands, Ghana, and Malaysia—representing Europe, Africa, and Asia, respectively. This diverse selection allows us to examine whether continental differences influence the interpretation and effectiveness of digital privacy laws. We further validate our choice of these countries in section~\ref{sec:problem}

\textbf{Digital Privacy Regulations in the Netherlands - }
The General Data Protection Regulation (GDPR)~\cite{gdprWhatGDPR} is a comprehensive data protection law implemented by the European Union (EU) in May 2018, and is enforced in the Netherlands. Its primary aim is to give individuals more control over their personal data and harmonize data protection laws across the EU. For users, some key points of interest include the requirement for explicit consent before their data can be collected and processed, the right to access and transfer their data, and the right to request the deletion of their data when it is no longer necessary. Moreover, organizations must notify users of data breaches that pose a risk to their rights and freedoms within 72 hours. Non-compliance with GDPR can result in significant fines, up to 4\% of annual global turnover or €20 million~\cite{gdprFines}, whichever is higher. \\

\textbf{Digital Privacy Regulations in Ghana - }
Ghana's digital privacy law, governed by the Data Protection Act, 2012 \cite{ghanaPrivacyAct}, aims to protect individual privacy by ensuring that personal data processing aligns with fundamental privacy rights. The act mandates fair and lawful data processing, data quality, and security requirements, and grants users rights to access, correct inaccuracies, and object to data processing under specific conditions. Explicit consent is required before collecting and processing personal data, similar to the GDPR. Additionally, the act establishes the Data Protection Commission\cite{dataProtectionCommissionGH}, an independent body that is responsible of compliance and enforces penalties for non-compliance with the law. \\

\textbf{Digital Privacy Regulations in Malaysia - }
In Malaysia, the Personal Data Protection Act (PDPA)~\cite{malaysiaPrivacyAct} established in 2010 regulates the processing of personal data in commercial transactions. The PDPA aims to safeguard individual privacy and ensure data is managed responsibly. Key aspects for users include principles of data processing that ensure data integrity, security, and lawful processing. Users have the right to access, correct, and withdraw consent for the use of their data. Organizations must obtain user consent before data collection and notify users about the purpose of data processing. Additionally, the PDPA sets specific conditions for transferring personal data outside Malaysia, ensuring it is protected abroad. Not complying with these regulations can lead to fines from the government.

\subsection{Comparing privacy regulations}
All three regulations emphasize giving users control over their personal data, providing rights to access, correct, and delete their data, which enhances their control over personal information. Transparency is another common aspect, with regulations requiring organizations to be clear about data processing activities, enhancing trust among users. Robust security measures are required to protect user data from breaches and misuse. Moreover, all three regulations include mechanisms for users to seek compensation in case of data protection violations. Despite these similarities, there are differences between the regulations. GDPR applies broadly to any organization processing data of EU residents, regardless of where the organization is located. In contrast, Ghana’s Data Protection Act primarily focuses on data processed within Ghana, while Malaysia’s PDPA mostly targets commercial transactions with specific conditions for international data transfers. Enforcement is also different. GDPR is enforced by data protection authorities in each EU member state, with substantial fines for non-compliance. In Ghana, the Data Protection Commission~\cite{dataProtectionCommissionGH} manages compliance and has the authority to enforce penalties. Malaysia’s enforcement is managed by the Department of Personal Data Protection~\cite{departmentDataProtectionMY}, which can also impose specific penalties for non-compliance.

\newpage

\section{Literature review} \label{sec:literature}

\subsection{Global variations in digital privacy laws}
Digital privacy laws protect the digital society and focus on different aspects, as we saw in the previous section. There are various digital privacy laws in the world that vary among each other. 
For instance, GDPR is more focused on where the business is established and PIPL is more focused on where the information processing happens \cite{chinabriefingPIPLGDPR}. The main aspect of DPP~\cite{rwandaDataLaw} is to empower citizens to protect the consumption of their non-consented data against third party agencies~\cite{rwandaDataLaw}.

There are discrepancies in privacy policies in the US and EU, and differences in the countries’ values, social norms, and interests result in a variance in regulations. Movius et al.~\cite{movius2009us} analyze the example of passenger name records in the travel industry as a case study and reach to the conclusion that US authorities are increasingly in favor of security, while European policy makers continue to emphasize personal freedoms. The author argues that due to the exchange of extensive volume of data between the US and EU, it is imperative for the economy and the protection of privacy rights of citizens to address the contrasting approaches on information privacy standards. 
There is further research comparing privacy regulations among different countries~\cite{baumer2004internet, custers2018comparison, movius2009us}. 

\subsection{Impact of GDPR on businesses and user behavior}
Since its implementation, GDPR has changed business practices and user behavior \cite{schmitt2020impact}. 
There is research~\cite{van2019does, layton2017gdpr, gruschka2018privacy} on analyzing the GDPR on how much it complies with data protection practices and it's implications on user behavior.
Oijen et al.~\cite{van2019does} examines how well GDPR helps people control their personal data. It finds that even though the GDPR includes rules like needing clear consent and allowing people to access, correct, move, and delete their data, people's behaviors often make these rules less effective. Problems like too much information leading to quick, uninformed consent, people choosing convenience over privacy, and feasible default settings that favor less privacy weaken GDPR's impact. The paper suggests that for the GDPR to work better, we need simpler privacy processes and more attention to how people actually behave.

Layton et al.~\cite{layton2017gdpr} compares GDPR with best practices in data protection and examines whether the GDPR aligns with the European Union's research and best practices. GDPR aims to give users control of their data and facilitate business operations. There is a gap, however, between ENISA's~\cite{enisaDesign} inputs for maximizing privacy and the GDPR's provisions. The GDPR focuses on specific regulations, institutions, and business practices but lacks discussion on improving user knowledge of privacy. The paper questions the assumption of digital literacy by GDPR and suggests that improving privacy, accountability, and trust through user behaviors could reduce costly compliance requirements. 

\subsection{Digital privacy practices in non-GDPR countries}
There exists research on digital privacy practices in the non-GDPR countries, namely Rwanda~\cite{mutimukwe2019information, nsengimana2017reflections} and Ghana~\cite{agyei2013empirical}.
Mutimukwe et al.~\cite{mutimukwe2019information} investigates information privacy protection (IPP) practices in Rwandan e-government services using international privacy principles as benchmarks. Their case study, involving three organizations, revealed that none fully complied with these principles, indicating concerns regarding the effectiveness of existing IPP practices. The absence of dedicated privacy policies and anticipated national regulations led to confusion in the e-government services, while inadequate and misleading practices undermined user control and accountability. The study emphasizes the necessity for coordinated efforts among government entities to raise awareness and enforce existing privacy laws, thus improving information privacy protection and enhancing user trust.
Nsengimana et al.~\cite{nsengimana2017reflections} discusses the impact of the Information and Communication Technology (ICT) revolution on personal privacy and its implications for the society, focusing on Rwanda's approach on protecting its citizens' privacy in the context of its digital transformation during the development of ICT in Africa, and its commitment to safeguarding personal privacy as a societal value.

There are certain limitations to privacy laws. Coleman et al.~\cite{coleman2018digital} examine digital colonialism as Western technology companies expand their presence in resource-rich, infrastructure-poor African countries, where data protection laws and regulations are not uniformly applied compared to Western standards. The paper analyzes the limitations of data protection laws, such as Kenya's 2018 data protection bill and the EU's General Data Protection Regulation (GDPR), in preventing digital colonialism.

\subsection{Effects of digital privacy on user trust}
Privacy regulations play a crucial role in shaping user trust and behavior in digital environments. Lancelot Miltgen and Smith~\cite{miltgen2015exploring} examines the relationship between information privacy regulations, perceived risks, trust, and user behavior. The study highlights how individuals' awareness of privacy regulations influences their perception of protection, which in turn builds trust in organizations and reduces concerns about privacy risks. Despite regulatory protections, users may still share personal information if they perceive significant benefits from doing so. The research suggests that understanding the balance between perceived risks and rewards is crucial for designing effective privacy regulations that foster trust and encourage responsible user behavior.

Privacy positively impacts both trust and ease of use in digital banking. Specifically, the more secure and protected users feel their personal information is, the more they trust the digital banking platform and find it easy to use~\cite{martinez2023does}. Similarly, effective privacy management and compliance with regulations can significantly improve user trust in digital platforms~\cite{quach2022digital}.

Aldboush et al.~\cite{aldboush2023building} focus on the fintech sector, analyzing how privacy regulations and ethical data practices affect customer trust. The study highlights the importance of corporate digital responsibility and compliance with data-protection laws to enhance trust. Transparent and responsible data handling practices are crucial for building and maintaining user trust in the digital finance industry. This aligns with findings by Kira~\cite{kira2021regulating}, who explores the impact of GDPR on user trust and organizational behavior. Kira's research indicates that strong privacy regulations like GDPR can enhance user trust by ensuring better protection of personal data and greater transparency from organizations.

Cao et al.~\cite{cao2022improving} explore how enhanced data privacy protections, such as those provided by the GDPR, impact consumer trust. Transparent privacy policies and explicit consent mechanisms increase consumers' perceptions of control over their data, thereby enhancing trust in e-commerce platforms. Firms adhering to strict privacy regulations benefit from increased customer loyalty and trust.  Similarly, Fox et al.~\cite{fox2022enhancing} investigates the impact of GDPR-compliant privacy labels on user perceptions of privacy risk, control, and overall trust in e-commerce vendors. A GDPR label is a proposed standardized label designed to provide clear and concise information about a company's data protection and privacy practices. The label aims to improve transparency and enhance consumer trust by making privacy practices easily understandable. The research shows that such a label could significantly boost consumer confidence in how their personal data is handled.

In the banking sector, Lappeman et al.~\cite{lappeman2023trust} examine the use of AI-driven chatbot services and how privacy concerns affect user trust and willingness to disclose personal information. The study concludes that robust privacy regulations and transparent data management practices are critical for building and maintaining user trust in digital services.

Overall, these studies collectively illustrate that privacy regulations are pivotal in enhancing user trust. By ensuring transparency, control, and security of personal data, these regulations help build a trustworthy digital environment, encouraging positive user behaviors and regulatory support.
\newpage
\section{Problem \& Research questions} \label{sec:problem}
Despite the existence of privacy regulations, digital services often fail to implement effective privacy practices, leaving users feeling unsafe and unprotected. 
There is a lack of research on the impact of digital privacy regulations and practices on user trust with the combination of comparing different countries. This study aims to address this gap by examining the impact of digital privacy regulations on user trust in the Netherlands, Ghana, and Malaysia. We chose these countries because we wanted to compare the distinct privacy regulations from countries in three different continents. Our contacts, resources and time-frame provided for this thesis led us to pick these three countries.
Due to having interviews from three different countries, this analysis will provide valuable insights into the global dynamics of privacy regulation and its implications for user trust across different continents.
We want to compare and understand how users perceive digital privacy regulations and what factors influence their trust in digital services. By exploring user perspectives in these three countries, this research can shed light on how digital privacy regulations affect users' sense of safety and privacy when engaging with digital services.
Therefore, the following research question (RQ) arises: \\
\textit{"How does the digital privacy regulations in the Netherlands, Ghana, and Malaysia impact user trust when consuming digital services?" }
From the RQ above, we derive the following sub-research questions (SRQ): \\
SRQ1: \textit{How does the digital privacy regulations in the Netherlands, Ghana and Malaysia impact photo sharing in social media services?} \\
SRQ2: \textit{How does the digital privacy regulations in the Netherlands, Ghana and Malaysia impact users' trust in businesses?} \\
SRQ3: \textit{How does digital privacy regulations in the Netherlands, Ghana, and Malaysia impact the users' trust in e-government services?}

\newpage
\section{Methodology} \label{sec:methodology}
We conduct exploratory and comparative multiple-case study to investigate the impact of digital privacy laws on user trust in digital services in the Netherlands, Ghana, and Malaysia. 
Our study examines how these regulations influence user trust in photo sharing, businesses, and e-government services. The comparative case study method is well-suited for this research because it allows for an in-depth understanding of how different regulatory environments impact user trust in digital services across diverse cultural and legal landscapes. By examining multiple cases, we can identify patterns and variations that a single case study might overlook. This approach provides a richer understanding of the subject matter, enabling us to draw more generalizable conclusions about the effectiveness of digital privacy regulations.

\subsection{Research design}
We interview 2 individuals each from the Netherlands, Ghana, and Malaysia to understand their perspectives on digital privacy. Our interviewees include digital privacy law experts, IT educators, and consumers of digital services. The qualitative data collected is analyzed within each specific context to grasp the dynamics of digital privacy regulations in different countries. With the comparative case study method, we focus on people who either consume digital services or posses extensive knowledge of digital privacy practices, which allows us to explore the similarities and differences between cases and later on propose a recommendation framework to improve the effectiveness of privacy practices. 

\subsection{Data collection}
To answer the research question, we conduct semi-structured interviews because it is more suitable for qualitative studies~\cite{bryman2016social} and we wanted to have some room for exploring different topics related to digital privacy during the interviews. 
We take two individuals from each country, totaling to six participants. Some interviews are conducted in-person, while others are done online. Interviewees are selected based on their expertise and our contacts. All data collection has been done according to ethical standards. Interviewee personal data are kept confidential at their request. 
A detailed list of interviewees and information about interviews is provided in Table \ref{table:interviews}. More information about interview content can be found here~\cite{interviewsLink}. The interviews transcripts and recordings are kept private in a repository and available at request.

\begin{table}[H] 
    \centering
    \begin{tabular}{ |c|c|c|c|c| }
    \hline
    Name & Organization & Country & Position & Duration \\ 
     \hline
     Interviewee (I1) & VU Amsterdam & Netherlands & Associate Professor & 57 minutes \\
     Interviewee (I2) & Freelance & Netherlands & ICT for development & 53 minutes \\
     Interviewee (I3) & VU Amsterdam & Ghana & PhD student & 42 minutes \\
     Interviewee (I4) & VU Amsterdam & Ghana & PhD student & 34 minutes \\
     Interviewee (I5) & UNIMAS & Malaysia & Senior lecturer & 30 minutes \\
     Interviewee (I6) & UNIMAS & Malaysia & Lecturer & 20 minutes \\
     \hline
    \end{tabular}

\caption{Metadata of each interviewee}
\label{table:interviews}
\end{table}

\subsection{Data analysis}
Data is collected using an audio recorder and then manually transcribed to create a document for each interview. Each case is analyzed separately to identify patterns and understand user perspectives on digital privacy. We utilize the phenomenological analysis method~\cite{lin2013revealing}, which focuses on uncovering the essence of users' experiences with digital privacy regulations. This approach is well-suited for prioritizing participants' perspectives, providing rich insights into how users perceive and interact with these regulations. The phenomenological approach is conducted using an exploratory multi-case study as follows: \\
\textbf{Data collection -} We conduct semi-structured interviews to gather rich data with detailed perspectives of participants' lived experiences in the area of digital privacy. \\
\textbf{Phenomenological bracketing -}
Before analyzing the data, we engage in phenomenological bracketing. This involves setting aside preconceptions, biases, and assumptions about the phenomenon of digital privacy to approach the data with openness and attentiveness to understand the participants' experiences as they describe them. \\
\textbf{Data analysis -}
We analyze each interview separately, focusing on identifying common themes and patterns related to participants' lived experiences about digital privacy. We seek to understand the underlying meanings from the user perspectives and perception on digital privacy. After analyzing our interviews, we come up with phenomenological aspects that are inline with our research questions, namely the general perspective, photo sharing, businesses, and e-government services. \\
\textbf{Cross-case analysis -}
We compare the analysis of interviews with each other, looking for similarities and differences in participants' lived experiences within different contexts in the realm of digital privacy. \\
\textbf{Interpretation and reporting -}
We interpret the findings, providing insights into the deeper meaning and significance of participants' lived experiences within the context of our research question. Lastly, we propose a recommendation framework consisting of sub-frameworks for the digital privacy regulations and practices.

\newpage

\section{Interview analysis} \label{sec:result}
In this section, we explore the perspectives of interviewees from the Netherlands, Ghana, and Malaysia, addressing each research question for each country. By presenting the observations in bullet points, we aim to provide a clear and precise overview of user viewpoints. It is important to note that the claims and opinions expressed are exclusively those of the interviewees, capturing their individual experiences and insights regarding digital privacy within their respective regions. 

\subsection{Netherlands}
We analyze the experiences of the interviewees from the Netherlands. This section has more content compared to other countries because we have more data related to GDPR due to interviews simply lasting longer than other countries. \\

\textbf{Privacy policies - } 
The interviewees from the Netherlands generally tend to trust the privacy policies or terms and conditions for commercial digital services (e-commerce) at first glance. The privacy policy indicates that they value the privacy of their users, so we tend to trust them. After starting to read the privacy policy, we understand that the policy is quite broad and vague. The policy mentions things like improving your user experience or sharing your data with selected business partners from the digital service. This is not reassuring at all for users of the service. 
Moreover, users tend to trust online digital services because their policies generally state that they value their users privacy. Hence, users trust the services without further reading the privacy policies. Nonetheless, the privacy policies of e-commerce services tend to be vague. 

Users are not given a choice when using certain digital services. For instance, a user might need to use a service their school or work is using, and that same user has to agree to the privacy policies of the service, because the user has to use the service. Hence, users agree to privacy policies without actually consenting to it because they have to use the digital services for whatever reason.
Nevertheless, one of the main goals of GDPR is to give control of user data to the users themselves~\cite{gdprWhatGDPR}. If the faculty or the affiliated company does not provide an option to the user regarding the usage of online digital services, it implicitly restricts the user's freedom of choice.

~\textit{~\textbf{Impact of digital privacy regulations - }}
GDPR is regarded as the most strictest and well-implemented privacy regulation globally, but its effectiveness in making the average citizen feel safer or more protected is limited. The primary privacy issues, such as tracking and digital advertising, are not fully addressed by GDPR. Despite these regulations, enforcement of GDPR isn't fully done due to understaffed agencies, which allows major privacy violators to bypass compliance while small businesses and organizations struggle with the administrative burden and fear of violating the law because of huge fines~\cite{gdprFines}. An anecdote a dutch interviewee highlights is about the practical challenges of GDPR, where a simple request for a neighbor's phone number was denied due to privacy concerns, illustrating a small consequence and burden on GDPR compliance.

Nonetheless, GDPR has made somewhat positive impact, such as compelling multinational companies to store EU data within Europe. The average citizen, however, might not fully exercise their rights under GDPR, and the regulation is perceived more as a starting point than a comprehensive solution.

\begin{tcolorbox}
\begin{description}
\itemsep-0.2em
    \mainfinding{analysis:netherlands:P1}{GDPR is a good starting point for privacy regulations and has some positive impact. The average citizen, however, does not feel safer or secure because of GDPR.}
\end{description}
\end{tcolorbox}

\begin{tcolorbox}
\begin{description}
\itemsep-0.2em
    \mainfinding{analysis:netherlands:P2}{Privacy policies of commercial digital services don't give enough trust to the users that their data is being protected properly.}
    \mainfinding{analysis:netherlands:P3}{Users give involuntary consent to digital services they are required to use.}
\end{description}
\end{tcolorbox}

\textbf{Balance between privacy and economy - }
The purpose of GDPR is two fold. The first purpose is that the European Union (EU) wants to protect the data protection interests of all the data subjects. The second purpose is to not limit the exchange of personal data within the European union. Hence, the purpose is both to protect the data protection rights of citizens and the other is to make sure that personal data flows freely within the European Union for both governmental and commercial institutions. Nobody talks about article 1 subsection 3 of the GDPR~\cite{gdprArt1}. 
The European Union is based on free flow of humans, capital and labor. Data is the new oil, and the EU wants to capitalize on data for Big Data purposes, training AI, and for other purposes. The EU doesn't want data protection to get in the way of economic growth. Hence, the aim for the EU is to balance data protection and economical growth. Moreover, the flow of personal data within the EU is well-protected by GDPR. 

\begin{tcolorbox}
\begin{description}
\itemsep-0.2em
    \mainfinding{analysis:netherlands:P4}{GDPR is trying to balance between data protection and economical growth.}
\end{description}
\end{tcolorbox}

There is a complex interaction between GDPR and the commercial interests of social media platforms. While social media companies benefit from users sharing extensive content to increase engagement and profitability, they simultaneously face regulatory requirements to protect user privacy.
The strategy employed by social media platforms to navigate this tension is about offloading the responsibility of data privacy to the users. Specifically, these companies transfer the responsibility for obtaining consent for shared content to the users themselves through their terms and conditions. This approach allows platforms to maintain high levels of user-generated content, which is essential to their business model, while simultaneously complying with GDPR.
Despite GDPR's intentions to enhance user privacy and control over personal data, the implementation and enforcement of these regulations in the context of social media are complicated by the platforms' underlying business models. This outsourcing of responsibility to users and the broad licensing terms for user content indicate that the influence of GDPR on social media practices may be limited, prompting further investigation into the efficiency of current regulatory frameworks in terms of truly safeguarding user privacy.

\begin{tcolorbox}
\begin{description}
\itemsep-0.2em
    \mainfinding{analysis:netherlands:P5}{Some social media companies offload privacy responsibilities to its users, freeing themselves from certain data privacy liabilities.}
\end{description}
\end{tcolorbox}

\textit{\textbf{Photo-sharing - }}
It is our freedom of expression right to share what happens in our lives, and if that is taking a photo on the public road, then let that be it. As long as the person in the photo is relatively anonymous and the photo is relatively by accident, then maybe it won't be a ~\textit{big deal} in terms of privacy invasion. If one tags a person in the photo, however, then you make them identifiable. Hence, the photo contains personal data of another person, invading privacy. Untagged photos are also personal data but it is harder to identify the untagged person's identity, unless one uses facial recognition software. Facial recognition software challenges the idea of anonymity in public places. There is a blurry relationship between data protection and freedom of expression.
The authorship of the photographer can be limited by the reasonable interest of the person in the photo. 
In contrast, for governments who require that you upload a photo the situation is different. The government has legal basis for asking our photo. We citizens hope that the government doesn't sell our photo or private data for other privacy invading purposes. 

\begin{tcolorbox}
\begin{description}
\itemsep-0.2em
    \mainfinding{analysis:netherlands:P6}{Freedom of expression can be limited by the content of the photo.}
\end{description}
\end{tcolorbox}

\textit{\textbf{Businesses - }}
The implementation of GDPR is positive, and has led to an improvement in transparency, with many businesses now disclosing their data processing practices on their websites. Users, however, find it challenging to verify the accuracy of these practices and privacy policies because of transparency and complexity issues. The compliance to privacy policies is only tested during certain issues like data breaches or company investigations, revealing false claims. Moreover, despite the necessity of using local e-commerce services, there is a lack of trust due to incidents like platform hacks that lead to anonymous calls. The reason is that small companies, like some local e-commerce, face challenges in securing data due to limited resources, require support mechanisms to aid their compliance efforts in digital privacy regulations. 

Businesses must have a legal basis, typically consent, to share data, specifying the purposes beforehand, leading many companies to hire consultants to ensure compliance. This has created a lucrative market for consultancy services. GDPR, which replaced the Data Protection Directive ~\cite{gdprArt94}, introduced stricter enforcement and higher fines, causing concern among companies despite the similarity in rules to the previous directive. Large corporations can afford legal experts to comply with or bypass GDPR, while small businesses often mimic competitors' privacy policies or risk non-compliance due to limited resources. Although large tech companies often get fined, they remain powerful and influential, which indicates that legal measures alone might be insufficient to control their impact on privacy. Enforcement of GDPR in the Netherlands is insufficient, with understaffing in government agencies responsible for tracking compliance, leaving many issues unresolved. While GDPR aims to protect personal data, its effectiveness is questioned due to these enforcement challenges.

\begin{tcolorbox}
\begin{description}
\itemsep-0.2em
    \mainfinding{analysis:netherlands:P7}{GDPR improved transparency in businesses regarding data privacy. There is a lack of trust in local e-commerce services due to hacking incidents.}
    \mainfinding{analysis:netherlands:P8}{Compliance to GDPR is a burden for small businesses. Large companies can use their resources to bypass GDPR.}
    \mainfinding{analysis:netherlands:P9}{Government is understaffed to track compliance in GDPR.}
\end{description}
\end{tcolorbox}

\textit{\textbf{E-government services - }}
The impact of the GDPR on privacy practices in e-government services is different than commercial services because governments can often justify data processing on the grounds of public interest. Governments process and share huge amounts of personal data among various agencies, sometimes leading to profiling and biased practices. For example, higher police surveillance in certain neighborhoods can create self-fulfilling prophecies, disproportionately targeting specific ethnic groups like Moroccan youths. Unlike commercial digital services, GDPR's "right to be forgotten" does not apply to government data needed for public services. 

Moreover, the impact of GDPR on privacy practices in Dutch e-government services increased government's focus on privacy. A significant issue, however, is still present, which is the reliance on US-based cloud providers like Azure, which can potentially provide data to the US government if requested. This poses potential privacy risks for users. While GDPR has had a positive impact on Dutch e-government privacy practices, complete trust in these services is challenging due to unavoidable data sharing among government departments for various purposes, which GDPR cannot fully prevent. 
Lastly, to improve GDPR, a public debate on its practical implications and societal values can be conducted making digital privacy more understandable for citizens, which can potentially lead to legal improvements. Additionally, the negative effects of data harvesting and addiction to social media, though not exact privacy issues, need legal attention as they are related to the broader impact of data harvesting.

\begin{tcolorbox}
\begin{description}
\itemsep-0.2em
    \mainfinding{analysis:netherlands:P10}{GDPR cannot prevent the consequences of relying on US-based cloud providers.}
    \mainfinding{analysis:netherlands:P11}{GDPR's "right to be forgotten" does not apply to the Dutch government.}
\end{description}
\end{tcolorbox}

\subsection{Ghana}
In this section we analyze the experiences of the two interviewees from Ghana, 
working as PhD students.
As mentioned before, the privacy law in Ghana is the data protection act established in 2012 by the National Information Technology Agency~\cite{ghanaPrivacyAct}.

\textbf{Digital privacy regulation - }
The data protection act prevents handling of private information without user consent. For instance, just like GDPR, the data protection act also prohibits people from giving out personal phone numbers without consent, which could lead to negative consequences. 
Nonetheless, majority of the society is not aware and do not adhere to the data protection act. And most of the time the law isn't practiced because the people that need to make sure that the law is being practiced are understaffed within the Ghanaian government.
Hence, people only become aware of the existence of the act if something bad, like being hacked, happens to them. 
People get fined or punished and only then learn about the data protection act because they don’t want to get fined or punished again.   
So people are only aware that digital privacy is a thing in Ghana if the law is practiced on them.
Hence, the data protection act is more of a silent law that the government and corporate organizations are aware of and utilize.

The widespread lack of awareness regarding digital privacy regulations in Ghana primarily comes from inadequate education concerning these new regulations. While numerous technological laws have been enacted in response to advancements in technology, they often remain enclosed within legal texts without being effectively communicated to the public. As a result, many individuals are unaware of these new regulations. This lack of knowledge is largely due to the absence of comprehensive educational initiatives aimed at explaining the laws in a manner that is accessible and understandable to the general public. 
Currently, the awareness of these digital privacy regulations is mostly limited to legal professionals, law students, or Information Technology professionals. Furthermore, the data protection act until recently was not that dominant in Ghana as compared to Europe. Thus people were more concerned about consuming or using the platforms but they were not much aware of what digital privacy protection is all about. 

The Data Protection Act has minimal impact on individuals' behavior when using digital services, especially for people who are unaware of the law. Those with a background in computer science, who are already aware of potential risks regardless of the law, are cautious about what information they share online. Despite the existence of data protection laws in Ghana, their implementation and enforcement are often insufficient, leading individuals to rely more on their own self-protective measures rather than the data protection act. While the laws are recognized as important for governing the use of others' data, people in Ghana do not always remember specific legal details because it's quite difficult and cumbersome to read long pages of law documents. Moreover, the sense of security in using digital services has remained unchanged before and after the enactment of the Data Protection Act, indicating that the law did not significantly alter perceptions of data privacy.

\begin{tcolorbox}
\begin{description}
\itemsep-0.2em
    \mainfinding{analysis:ghana:P12}{Most people in Ghana only become aware of digital privacy regulations after a violation impacts them directly.}
    \mainfinding{analysis:ghana:P13}{The Data Protection Act has minimal impact on people's behavior in Ghana, as insufficient enforcement leads privacy-aware people to protect their digital privacy on their own.}
\end{description}
\end{tcolorbox}

\textit{\textbf{Photo sharing - }}
The Data Protection Act affects photo sharing on social media in various ways. Bloggers and careful users often add disclaimers when sharing photos. Many are also cautious because of internet fraud and not knowing much about digital safety. 
They're more comfortable sharing photos on platforms like WhatsApp where they can control who sees them. Nevertheless, most people don't understand the risks of sharing personal information online, and are not aware of the Data Protection Act that's supposed to protect them.
People are more open with family and friends but more careful with others.
For instance, friends usually don't mind being tagged in pictures, so permission isn't often sought. Sharing photos of strangers without their consent, however, can cause problems, as people value their privacy and might confront you. Moreover, the law allows people to sue if their privacy is violated, but they must actively pursue this. The reason for taking and sharing the photo is also important. If it's done with good intent, it’s usually acceptable, but secret photos can lead to privacy issues that might not be addressed.
Moreover, people don't always realize what they share online can reach a wider audience. This misunderstanding comes from how social media used to be more private, but now it's more public, and many people in Ghana haven't caught up with this change.

\begin{tcolorbox}
\begin{description}
\itemsep-0.2em
    \mainfinding{analysis:ghana:P14}{Awareness regarding the digital privacy rights effects the users' behaviours in social media services.}
    \mainfinding{analysis:ghana:P14.2}{People are not fully aware of the limits of social media when they share something.}
\end{description}
\end{tcolorbox}

\textit{\textbf{Businesses - }}
People in the IT and business sector are probably more aware of the laws. The data protection act is probably enforced more on companies and businesses because they handle vast amount of data. 
Businesses are aware of the law and are careful about it. 
Data sharing between companies sometimes happens. For example, telecommunication companies share phone numbers with telemarketing companies. 
Nonetheless, there's a significant problem with illegal data mining in Ghana, where companies obtain personal information without consent and sell it to others. This results in individuals receiving messages from unknown numbers and raises concerns about privacy violations. While the Data Protection Act should protect against such practices and impose fines on violators, government enforcement is lacking because of staffing and infrastructure limitations. Without sufficient resources to monitor and fine companies for breaching digital privacy laws, protecting the privacy rights of the society can be an issue.

Moreover, the society hopes that most of the companies adhere to the digital privacy laws. Most of the IT companies in Ghana are startups. Most of them are cautious regarding digital privacy because they wanna succeed and don’t want to take any action that is against the law. Those who may bypass the law might be huge companies. Furthermore, most Ghanaians would trust their local companies more than the international ones. When it comes to digital privacy, most of the concern is with international companies rather than local companies. This might be because people have used local business services for a long time and nothing has gone wrong.
Local businesses are often trusted more than international ones, and there's a belief that the government may punish local businesses for privacy violations but may be less strict with multinational corporations due to tax revenue considerations. Companies with limited financial resources often hire contract workers to ensure compliance with privacy regulations.

\begin{tcolorbox}
\begin{description}
\itemsep-0.2em
    \mainfinding{analysis:ghana:P15}{Government is understaffed to punish companies violating digital privacy regulations.}
    \mainfinding{analysis:ghana:P16}{The track record of local companies lead to people trusting them more than international companies.}
\end{description}
\end{tcolorbox}

\textbf{\textit{E-government services}}
Data protection is the top policy that is being practices by the government when implementing e-government services.
There is a significant level of trust in e-services, particularly in e-government platforms handling personal and financial services, perhaps because the government adheres to digital privacy practices and there have been no notable incidents of data breaches. Organizations with expertise in data privacy, such as National Information Technology Agency (NITA) and the National Communication Agency (NCA), play a crucial role in fostering this trust. NITA, which consists of a mix of IT industry experts and government officials, contributes to a balanced approach to data privacy. The Data Protection Act and is an example of NITA’s work~\cite{ghanaPrivacyAct}. This balance between industry and government personnel leads to greater confidence in e-government services.

The trust in the government's digital privacy efforts, despite suspicions in other areas, can be attributed to the IT revolution. The government's collaboration with industry experts to develop IT infrastructure has led to a different dynamic in the IT field compared to other government sectors. The involvement of trusted IT professionals within the government has resulted in a distinctive approach to IT policies in several African countries, including Ghana. The responsiveness of organizations like the NCA to public concerns, as demonstrated by their reversal of the decision regarding Starlink~\cite{ghanaStarlink}, illustrates the strong influence and active participation of IT professionals in shaping governmental IT policies.

\begin{tcolorbox}
\begin{description}
\itemsep-0.2em
    \mainfinding{analysis:ghana:P17}{Trust in e-government services comes from the balance between industry and government officials.}
\end{description}
\end{tcolorbox}

\subsection{Malaysia}
In this section we analyze the experiences of the two interviewees from Malaysia who are academic faculty members in a university in Malaysia.
As mentioned before, the privacy law in Malaysia is the personal data protection act established in 2010 by the Malaysian government~\cite{malaysiaPrivacyAct}. \\

\textbf{Digital privacy regulation - }
In Malaysia, the personal data protection act exists but is not effectively enforced, with no significant penalties for breaches. This has led to skepticism about its impact on privacy and security when using digital services. The law, established in 2010, is currently under refinement, potentially due to its age and evolving digital landscape.

In practice, the protection of personal information largely depends on the platform's reputation and security measures. A popular e-commerce platform, for instance, is trusted because of its robust security features, such as two-factor authentication (2FA), and its lack of security incidents. 
This 2FA security measure is also employed for other applications, such as email logins and student accounts, providing a broader sense of safety. 
Hence, trust in digital services is based more on the brand's security practices, especially the 2FA security measure, than on the digital privacy laws in Malaysia.

\begin{tcolorbox}
\begin{description}
\itemsep-0.2em
    \mainfinding{analysis:malaysia:P18}{The trust in digital services is based on the digital brand's security practices and lack of security incidents, rather than the laws.}
\end{description}
\end{tcolorbox}

\textit{\textbf{Photo-sharing - }} 
The impact of privacy practices on photo-sharing activities on online digital platforms is insignificant. People often upload photos without seeking permission and only remove them if someone complains. There is little awareness of or concern for the personal data protection act (PDPA), and no fines or punishments are enforced for sharing unwanted photos online. The primary response to issues like this is to remove the photo after complaints. Respect for privacy in photo-sharing is driven more by individual ethics than by law. For example, in Malaysia, the government prohibits teachers from sharing photos of children under 13 on social media and prohibits children of this age group from having social media accounts. Due to ongoing refinements of the law, however, enforcement is lacking, and compliance with the law varies among individuals. Teachers often remind parents of this rule.

Photo-sharing activities in social media is not different than in other online digital platforms. Respect for privacy on social media is minimal, with most users uploading photos without considering consequences or privacy breaches. Understanding of digital privacy is limited. While Malaysians are tech-savvy and proficient in using digital technology, they lack awareness of the rules governing its use. Those who are aware of privacy regulations tend to follow them, often seeking consent before sharing someone else's private information. The lack of adherence to rules is attributed to a lack of awareness rather than a deliberate choice to ignore them. Government efforts to increase awareness through PDPA campaigns have not been very effective, as many find the rules too lengthy and complex to read.

On the other hand, despite digital privacy being understood as a human right, it is less respected than physical privacy. Digital privacy education is minimal, with limited theoretical education at the university level. There is a call for the government to enhance public awareness and education on digital privacy to simplify the understanding of complex laws.

\begin{tcolorbox}
\begin{description}
\itemsep-0.2em
    \mainfinding{analysis:malaysia:P19}{Despite digital literacy, Malaysians often disregard privacy practices in photo-sharing on online digital platforms due to a lack of awareness and ineffective enforcement of the personal data privacy act.}
\end{description}
\end{tcolorbox}

\textit{\textbf{Businesses - }}
In the context of data sharing among businesses, business-to-business (B2B) and business-to-consumer (B2C) transactions are generally perceived as secure, with minimal concerns about data leakage. Trust in these transactions is primarily based on the brand reputation and security practices of the platforms used, such as the implementation of two-factor authentication. Local and international companies with strong branding are trusted similarly, as seen with the e-commerce platform Shopee~\cite{shopeeLink}, which, despite being based in Singapore, is highly trusted by Malaysians. Trust to digital service is based on brand rather than the personal data protection act. Instances of data breaches or privacy violations in Malaysia are rare and often unproven. Nonetheless, customers of Cimb~\cite{CimbMY} claimed that there had been a data breach, yet the company Cimb denied such an incident~\cite{cimbBreach}. 
There is, however, a cautious approach towards sharing financial information with anyone and sharing sensitive information with international companies due to potential data theft concerns.

\begin{tcolorbox}
\begin{description}
\itemsep-0.2em
    \mainfinding{analysis:malaysia:P20}{Trust in data sharing among businesses in Malaysia is primarily based on brand reputation and security practices rather than the personal data protection act.}
\end{description}
\end{tcolorbox}

\textbf{\textit{E-government services - }}
Government services utilize privacy practices that are inline with the personal data protection act. 
For instance, the government has introduced a digital ID that is utilized in Malaysia's e-government services. This is a step towards more security and privacy when using e-government services, although concerns about data safety persist among some users. The digital ID, which facilitates Single-Sign-On (SSO) for accessing government services, is part of a broader transition from physical IDs. 

There is a trust in the government that they will protect the data of their citizens. There have been no reported incidents undermining trust in government data protection thus far. The relatively slow adoption of digital services is attributed to inefficiencies in service delivery when using digital ID, thus many prefer in-person interactions for governmental matters. Another aspect of trust in the government is that the Malaysian government is in the process of refining the personal data protection act, reflecting their growing awareness and the need for robust data protection in the context of advancing AI technologies and the shift towards a digital economy.

\begin{tcolorbox}
\begin{description}
\itemsep-0.2em
    \mainfinding{analysis:malaysia:P21}{Malaysians trust the government with their digital privacy. Yet, the digitalization efforts of the government in certain areas still need improving.}
\end{description}
\end{tcolorbox}

\subsection{Comparison between countries} \label{subsec:comparison}
This comparative analysis highlights how the effectiveness and perception of digital privacy regulations vary significantly across the Netherlands, Ghana, and Malaysia, influenced by factors such as enforcement, public awareness, and the robustness of individual platform privacy measures. A summary of the impact of privacy regulations is presented in table~\ref{tabel:summary_digital_reg}.

\begin{table}[H]
\centering
\begin{tabular}{|p{3.5cm}|p{3.5cm}|p{3.5cm}|p{3.5cm}|}
\hline
\textbf{Aspects} & \textbf{Netherlands} & \textbf{Ghana} & \textbf{Malaysia} \\
\hline
\textbf{General perspective} & 
    GDPR is a good starting point, average citizen does not feel safer. Privacy policies are still vague. EU is trying to balance between economical growth and digital privacy. Sometimes compliance to GDPR can be a burden & 
    Data Protection Act has minimal impact on people’s behavior in Ghana because most people are not aware of the act. Insufficient enforcement leads privacy-aware people to protect their digital privacy with their own ways & 
    Trust in digital services is based on the company's track record of security and privacy incidents. There is a refinement in the personal data protection act. \\
\hline
\textbf{Photo sharing} & 
    Freedom of expression can be limited by content & 
    Awareness regarding the digital privacy rights effects the users’ behaviours in
social media services & 
    There is no privacy awareness about the regulations and photo-sharing activities \\
\hline
\textbf{Businesses} & 
    Government is understaffed to keep track of compliance with GDPR. & 
    Government is understaffed to keep track of compliance to the privacy act. Local companies are trusted more than international ones because of track record. NITA and NCA balance IT and government expertise & 
    Trust in businesses, regardless of the origin of the company, comes from the track record of security and privacy incidents \\
\hline
\textbf{E-government} & 
    Challenges in trust, public debate needed & 
    There is a trust in e-government services which comes from the balance between industry and government officials in digital service policies & 
    Trust exists in the government, regardless most people go to government offices to handle their affairs. E-government services are not widely used. \\
\hline
\end{tabular}
\caption{Table containing the summary of the impact of digital privacy regulations}
\label{tabel:summary_digital_reg}
\end{table}

\textbf{\textit{Digital privacy regulations - }} 
General trust and awareness of digital privacy policies differ per country. In the Netherlands, there is initial trust in privacy policies. Upon closer inspection, however, this trust diminishes due to vagueness of the privacy policies. 
Ghana has a low general awareness of the Data Protection Act, and people only become aware of the act after an incident similar to Malaysia.
Similar to Ghana, people are not aware that there data privacy is protected by regulations in Malaysia. There is also skepticism due to lack of enforcement and trust is based more on platform reputation than in digital privacy laws.
The privacy acts in Ghana and Malaysia are not known by their citizens as Dutch citizens know the GDPR. So the awareness of GDPR is higher than the other two privacy acts in Ghana and Malaysia.

The effectiveness and implementation of each law differs per country. In the Netherlands, GDPR is seen as strict but is limited in addressing primary privacy concerns. Just like in Ghana and Malaysia, there are enforcement issues of the privacy regulations due to understaffing in the government. Nonetheless, the fines of GDPR are much higher than in Ghana and Malaysia, making it more compelling to comply with.

The practical impact and behavior of each privacy regulation differs per country. Complying with GDPR in the Netherlands causes practical challenges in daily interactions as people are afraid to be fined. 
In Ghana and Malaysia there is minimal change in behavior. People rely on self-protection due to lack of awareness and enforcement of the law. So trust is mostly based on individual platform security practices.

\textbf{\textit{Photo sharing - }}
In the Netherlands, there is a concern for the balance in freedom of expression and privacy in photo sharing activities, with emphasis on anonymity and consent. As mentioned before, GDPR is known but face practical enforcement challenges, especially with tagging and facial recognition issues.
Both in Ghana and Malaysia, photo-sharing practices are cautious among users that know the law. Nonetheless, regulations are not known well among the general public, and awareness is typically after an incident or fine. 
Hence, photo-sharing practices show minimal concern for privacy laws, with actions driven by individual ethics rather than legal mandates.
Regulations lack effective enforcement, and awareness campaigns by the government related to digital privacy are insufficient, leading to widespread non-compliance due to complexity of the law and a lack of understanding.

\textbf{\textit{Business - }}
Transparency and trust has improved in the Netherlands with GDPR, but there is still low trust in local e-commerce services due to data breaches. Privacy can be seen as a luxury, with wealthier people affording more secure services, like paying for certain software services or buying more expensive hardware that does not track the user. High compliance costs with GDPR lead to a market for consultancy services. Small businesses struggle with compliance, while large corporations manage but still face fines, and some actually get away with wrongdoings. Enforcement of the digital privacy regulations is lacking due to understaffing in the government.

In Ghana, there isn't even sufficient awareness among the general public regarding digital privacy, with businesses more aware of the law. Despite the law, there are still trust issues in businesses due to illegal data mining and lack of enforcement of the digital privacy law. Most local businesses and startups are careful about privacy due to the desire for success and adherence to the law to also not get fined while larger companies might bypass laws. Hence, local businesses are trusted more than international ones.
Unfortunately, just like the other two countries, enforcement of the digital privacy regulations is lacking due to understaffing in the government.

Moreover, in Malaysia, the trust in digital services is mostly based on brand reputation rather than legal compliance. There is not much concern regarding data breaches due to past reputations of businesses.
Compliance in digital privacy practices is mostly driven by individual ethics rather than the local digital privacy law.
Like The Netherlands and Ghana, enforcement of the digital privacy regulations is lacking due to understaffing in the government.

\textbf{\textit{E-government services - }}
The Dutch government justifies processing of personal data on public interest grounds, leading to extensive data sharing among governmental departments and potential profiling and bias. 
Public debate and awareness are needed to improve trust and understanding of GDPR's implications on e-government services.
GDPR has increased focus on privacy in the government, but challenges like reliance on US-based cloud providers persist. 

In Ghana, e-government services comply to the local digital privacy law, with no notable data breaches demolishing trust.
IT experts are involved in policy-making for digitalization practices, which leads to more trust from the society. Thus, organizations like NITA and NCA play crucial roles in data privacy, balancing IT industry and government expertise.

E-government services in Malaysia comply with the local digital privacy law, with measures like digital IDs for security.
There is trust in government data protection. Nonetheless, slow adoption of digital services due exist due to inefficiencies in governmental processes. The government is refining the the local digital privacy law, reflecting the need for more robust digital protection measures in the digital economy.

\newpage
\section{Recommendation framework} \label{sec:recommendation_frame}
Based on our interview analysis and comparison between countries, we propose a recommendation framework that could be applied to Netherlands, Ghana, and Malaysia to improve digital privacy practices. We construct our recommendation framework on principles for digital development~\cite{digitalDev} and academic research~\cite{schaar2010privacy, binns2018s, cavoukian2009privacy, acquisti2015privacy, ahmad2014information, martin2017role, lefkovitz2020nist} to ensure it is robust, evidence-based, and aligned with best practices in the field.
By integrating these principles and research, our recommendation framework benefits from a well-rounded foundation that addresses the technical, ethical, and practical dimensions of digital development and data protection. This approach ensures that our framework is not only theoretically solid but also practical and applicable in real-world settings.

We focus on the following digital development principles: \textit{create open and transparent practices }, \textit{establish people-first data practices}, and \textit{use evidence to improve outcomes}. 
Open and transparent practices means that having and maintaining trust in digital ecosystems. People have confidence in digital ecosystems which is established through open and transparent practices. For instance, these practices include but not be limited with the use of agile methodologies, open data, and open source software. We pick this principle because many privacy policies and regulations are too complex to comprehend, leading users to consent on policies they do not comprehend. Examples of not practicing this principle might lead to the following perspectives P-\ref{analysis:netherlands:P2} and P-\ref{analysis:netherlands:P3}.

Establish people-first data practices means prioritizing people's rights and needs when handling their data, ensuring that value is returned to the data subjects. This includes obtaining informed consent, adhering to data standards, and enabling users to control and benefit from their data. Violating these principles can lead to harm, such as data breaches or discrimination. Examples of not practicing this principle might lead to the following perspectives P-\ref{analysis:netherlands:P1}, P-\ref{analysis:netherlands:P2}, P-\ref{analysis:netherlands:P5}, P-\ref{analysis:netherlands:P7}, P-\ref{analysis:netherlands:P11}, and P-\ref{analysis:ghana:P16}

\begin{figure}[H]
    \centering
    \includegraphics[width=1\linewidth]{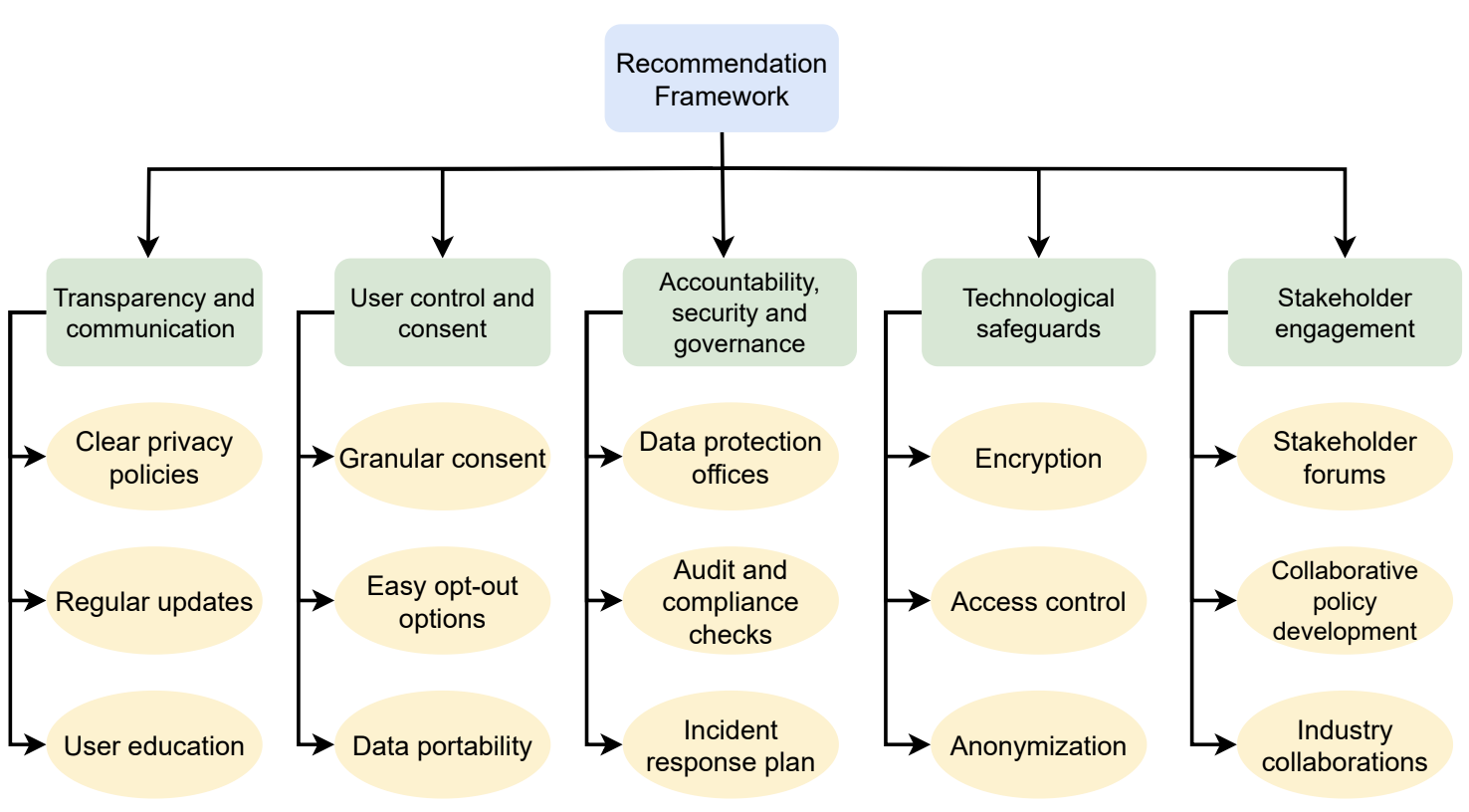}
    \caption{Recommendation framework broken down to its sub-frameworks and the components of those sub-frameworks. Green boxes represent sub-frameworks and yellow oval shaped figures represent the components.}
    \label{fig:recommendation-framework}
\end{figure}

Use evidence to improve outcomes means that impact depends on continuously gathering, analyzing, and utilizing feedback to understand the outcomes of digital services for people, using both technological and analogue methods. This holistic approach ensures that digital policies and solutions are continuously improved based on meaningful, people-centered metrics. Without this, initiatives like GDPR or the personal data protection act may achieve efficiency but fail to recognize or enhance their real impact on people and communities. Examples of not practicing this principle might lead to the following perspectives P-\ref{analysis:netherlands:P1}, P-\ref{analysis:netherlands:P8}, P-\ref{analysis:netherlands:P9}, P-\ref{analysis:ghana:P12}, P-\ref{analysis:ghana:P13}, P-\ref{analysis:ghana:P15}, P-\ref{analysis:malaysia:P18}, P-\ref{analysis:malaysia:P19}, P-\ref{analysis:malaysia:P20}, and P-\ref{analysis:malaysia:P21},  

Moreover, we come up with five sub-frameworks and visualize the general outlook of our recommendation framework in figure~\ref{fig:recommendation-framework}. Also, in table~\ref{tabel:recommendation_framework} we match the user perspectives with the sub-frameworks that could provide valuable improvements to the digital privacy practices per country.

\begin{table}[H]
\centering
\begin{tabular}{|p{3.5cm}|p{3.5cm}|p{3.5cm}|p{3.5cm}|}
\hline
\textbf{Sub-frameworks} & \textbf{Netherlands} & \textbf{Ghana} & \textbf{Malaysia} \\
\hline
\textbf{Transparency and communication} & 
    P-\ref{analysis:netherlands:P2} & 
    NA & 
    NA \\
\hline
\textbf{User control and consent} & 
    P-\ref{analysis:netherlands:P3} & 
    NA & 
    NA \\
\hline
\textbf{Accountability, security and governance} & 
    P-\ref{analysis:netherlands:P5}, P-\ref{analysis:netherlands:P9} & 
    P-\ref{analysis:ghana:P13}, P-\ref{analysis:ghana:P15}, P-\ref{analysis:ghana:P17} & 
    P-\ref{analysis:malaysia:P19}, P-\ref{analysis:malaysia:P21} \\
\hline
\textbf{Technological safeguard} & 
    P-\ref{analysis:netherlands:P7} & 
    P-\ref{analysis:ghana:P16} & 
    P-\ref{analysis:malaysia:P18}, P-\ref{analysis:malaysia:P20} \\
\hline
\textbf{Stakeholder engagement} &
    P-\ref{analysis:netherlands:P1}, P-\ref{analysis:netherlands:P4}, P-\ref{analysis:netherlands:P8}, P-\ref{analysis:netherlands:P9}, P-\ref{analysis:netherlands:P10} &
    P-\ref{analysis:ghana:P12}, P-\ref{analysis:ghana:P13}, P-\ref{analysis:ghana:P14}, P-\ref{analysis:ghana:P16}, P-\ref{analysis:ghana:P17} &
    P-\ref{analysis:malaysia:P19}, P-\ref{analysis:malaysia:P20} \\
\hline
\end{tabular}
\caption{Table matching the perspective of the users per country with the potential solutions involving the sub-frameworks of the recommendation framework}
\label{tabel:recommendation_framework}
\end{table}

\subsection{Transparency and communication framework}
This framework focuses on enhancing user trust through clear communication and transparency about data handling practices. The main references that are used when forming this framework are from Schaar et al.~\cite{schaar2010privacy}, Binns et al.~\cite{binns2018s}, and open and transparent practice principle from the digital development framework. Schaar discusses the concept of Privacy by Design, emphasizing the importance of integrating privacy considerations into the design and architecture of IT systems and business practices from the general outline. Binns explores user perceptions of algorithmic decision-making, emphasizing the need for transparency and clear communication to build trust and ensure fairness in data handling practices.
The components of this framework are:
\begin{itemize}
    \item \textbf{Clear privacy policies:} Ensure that privacy policies are written in clear, non-technical language that users can easily understand.
    \item \textbf{Regular updates:} Provide regular updates about any changes in privacy policies or data handling practices.
    \item \textbf{User education:} Implement educational programs to increase users' understanding of their privacy rights and the protections offered by regulations.
\end{itemize}

\subsection{User control and consent framework}
The user control and consent framework focuses on empowering users by giving them control over their data and ensuring that users know what they are giving consent for. The main references that are used when coming up with this framework are from Cavoukian et al.~\cite{cavoukian2009privacy}, Acquisti et al.~\cite{acquisti2015privacy}, and establish people first data practices from the digital divide framework. Cavoukian discusses the 7 foundational principles of privacy by design, advocating for user-centric controls and transparent data practices to empower users. Acquisti looks into the interaction between privacy, user behavior, and information systems, highlighting the importance of user control and informed consent in the digital age.
The components for this framework are:
\begin{itemize}
    \item \textbf{Granular consent:} Allow users to provide consent for specific data processing activities rather than a blanket consent for all activities.
    \item \textbf{Easy opt-out options:} Ensure users can easily opt out of data processing activities they do not agree with.
    \item \textbf{Data portability:} Provide users with the ability to easily transfer their data to other service providers.
\end{itemize}

\subsection{Accountability, security and governance Framework}
This framework is about building trust by demonstrating accountability and robust governance in data protection practices. It is based on the research of Ahmad et al.~\cite{ahmad2014information} and Martin et al.~\cite{martin2017role}. Ahmad explores various information security strategies employed by organizations, advocating for a comprehensive, multi-strategy approach. It emphasizes the integration of diverse security measures such as prevention, detection, and response to effectively protect information systems. The study highlights the need for organizations to adopt a comprehensive security framework that includes appointing security officers, conducting regular audits, and developing robust incident response plans. The findings indicate that a multi-level security strategy is essential for mitigating risks and ensuring the protection of organizational data.
Martin explores the critical role of data privacy in marketing, emphasizing its impact on consumer trust and business practices. He examines various theoretical perspectives and empirical findings on data privacy, addressing the psychological, societal, and economic dimensions. The authors argue that effective privacy management can enhance consumer trust and loyalty, proposing a robust governance framework for marketers to manage privacy concerns responsibly and ethically. In short the study highlights the necessity for transparency, accountability, and user-centric privacy controls to build and maintain trust.
Components for this framework are:
\begin{itemize}
    \item \textbf{Data protection officers (DPOs):} Appoint DPOs to oversee compliance with privacy regulations and handle user concerns.
    \item \textbf{Audit and compliance checks:} Conduct regular audits to ensure compliance with privacy regulations and best practices.
    \item \textbf{Incident response plan:} Develop and communicate a clear incident response plan for data breaches.
\end{itemize}

\subsection{Technological safeguards framework}
The Technological Safeguards Framework is about enhancing user trust by implementing strong technological safeguards to protect personal data. This framework is based on the NIST (National Institute of Standards and Technology)~\cite{lefkovitz2020nist}, where NIST provides guidelines for managing privacy risks through a structured approach, offering strategies for implementing strong privacy safeguards like encryption and access controls.
The components for this framework are:
\begin{itemize}
    \item \textbf{Encryption:} Use strong encryption methods to protect data at rest and in transit. 
    \item \textbf{Access controls:} Implement strict access controls to ensure that only authorized people can access sensitive data.
    \item \textbf{Anonymization and pseudonymization:} Use techniques to anonymize or pseudonymize data to protect user identities.
\end{itemize}

\subsection{Stakeholder engagement framework}
This framework is about fostering trust through active engagement with stakeholders, including users, regulatory bodies such as government agencies, and industry groups. The main reference for this framework is from the use evidence to improve outcomes principle from the digital development framework.
The main components of this framework are:
\begin{itemize}
    \item \textbf{Stakeholder forums:} Organize regular forums and gatherings to engage with users and gather feedback on digital privacy practices.
    \item \textbf{Collaborative policy development:} Involve users and other stakeholders in the development and refinement of digital privacy policies.
    \item \textbf{Industry collaboration:} Work with industry experts to develop, adjust, and adopt best practices in digital privacy protection.
\end{itemize}

\newpage
\section{Conclusion \& further discussion} \label{sec:conclusion}
This study delves into the impact of digital privacy regulations on user trust across three distinct regions: the Netherlands, Ghana, and Malaysia. The analysis shows how awareness, enforcement, and user behavior are influenced by cultural, economic, and regulatory factors.

In the Netherlands, the General Data Protection Regulation (GDPR) is widely recognized and has enforced significant changes in business practices and user behavior. The practical impact of GDPR, however, depends on many other situations. While it has improved transparency and given users more control over their personal data, challenges remain. Users often find privacy policies vague and complex, leading to a low level of compliance. Furthermore, enforcement issues, primarily due to understaffed regulatory governance bodies, limit the regulation's effectiveness. This means that while GDPR sets a high standard, its real-world impact can sometimes fall short of its actual intentions.

Ghana presents a contrasting scenario where the Data Protection Act is not as widely known or enforced as GDPR. Public awareness is significantly lower, and many citizens only become aware of the regulations after experiencing a privacy violation. The lack of enforcement and public education leads to a reliance on personal protective measures. The involvement of IT industry experts in government policy-making, nonetheless, leads to a certain amount of trust in e-government services, which is a positive outcome of the collaborative approach in Ghana's government.

Malaysia's experience with the Personal Data Protection Act (PDPA) emphasizes a similar lack of public awareness and effective enforcement like Ghana. The trust in digital services in Malaysia is more closely tied to the reputation and perceived security of individual platforms rather than the legal framework. The ongoing refinement of the PDPA indicates a growing recognition of the need to improve digital privacy protections in response to evolving digital systems.

This comparative analysis highlights several critical insights. Firstly, awareness and education are imperative. Higher levels of public understanding of privacy regulations, as seen in the Netherlands, correlate with greater trust and more informed user behavior. In contrast, the low awareness in Ghana and Malaysia diminishes the effectiveness of these laws. Secondly, enforcement is an important aspect of effective privacy regulation. The understaffed regulatory bodies in all three regions is an obstacle to successful execution of the privacy laws. Lastly, cultural and economic contexts play a crucial role. The localized adaptation of privacy regulations and the balance between governmental and industry expertise, particularly evident in Ghana, illustrate the importance of adapting privacy strategies to specific regional dynamics. 

\subsection{Contributions}
This study contributes to the broader understanding of digital privacy regulations and their impact on user trust in several ways. The comparative analysis in section 5.4 provides a robust method for evaluating the effectiveness of privacy laws across different countries and emphasizes the importance of considering local contexts when assessing regulatory impact. We also propose a recommendation framework, where each sub-framework could be used to improve a different aspect of digital privacy, aiming to improve the overall effectiveness of digital privacy regulations and practices. 

\subsection{Further discussion}
\textbf{Cross-Cultural implications of privacy regulations -} 
One area for further discussion is the cross-cultural implications of privacy regulations.
Cultural differences play a crucial role in how privacy laws are perceived and implemented. For example, the Netherlands, with its robust legal infrastructure and high public awareness, is different than Ghana and Malaysia, where cultural norms and lower awareness influence the effectiveness of privacy laws. Future research could look into how cultural values shape attitudes towards privacy and compliance, exploring whether culturally adapted privacy regulations could enhance effectiveness and user trust in different regions.

\textbf{Technological advancements and privacy -}
Technological advancements, such as artificial intelligence (AI) and machine learning, present both opportunities and challenges for digital privacy. While these technologies can improve data security and privacy management, they also create new risks and ethical dilemmas. For instance, the use of facial recognition technology raises significant privacy concerns. Further research could examine how emerging technologies intersect with privacy regulations, and how laws can evolve to address new challenges while leveraging technological benefits to enhance privacy protection.

\textbf{Effectiveness of enforcement mechanisms -}
The effectiveness of enforcement mechanisms is another critical area for further discussion. This study found that enforcement is a significant challenge across all regions examined, primarily due to understaffed regulatory bodies. Investigating alternative enforcement strategies could provide insights into more efficient and effective ways to enforce privacy regulations. Additionally, comparative studies on the enforcement models of different countries could identify best practices and innovative approaches to ensure compliance.

\bibliographystyle{abbrv}
\bibliography{main}

\end{document}